\documentclass[epj]{svjour}
\usepackage{graphics}
\usepackage{amsmath}
\usepackage[export]{adjustbox}
\usepackage{color}
\usepackage{graphicx}
\usepackage{dcolumn}
\usepackage{bm}
\usepackage{hyperref}
\usepackage{float}
\usepackage{cite}
\usepackage[caption=false]{subfig}
\usepackage{subfloat}
\usepackage{hyperref}
\usepackage{fancyhdr}
\hypersetup{
colorlinks=true,
linktoc=page,    
linkcolor=blue,
citecolor=blue,
urlcolor=blue,
colorlinks=true,
}

\newcommand{\be}{\begin{equation}}
\newcommand{\ee}{\end{equation}}
\newcommand{\bea}{\begin{eqnarray}}
\newcommand{\eea}{\end{eqnarray}}
\newcommand{\mM}{\mathcal{M}}
\newcommand{\mL}{\mathcal{L}}
\newcommand{\mD}{\mathcal{D}}

\newcommand{\pd}{\partial}
\newcommand{\nn}{\nonumber}

\newcommand{\dd}{\,\mathrm{d}}
\newcommand{\vpp}{\Psi}
\newcommand{\mS}{\mathcal{S}}

\newcommand{\A}{\text{Area}}

\begin{document}

\title{A black hole toy model with non-local and boundary modes from non-trivial boundary conditions}
\author{Peng Cheng\inst{1,2}}
\institute{
1) Center for Joint Quantum Studies and Department of Physics, School of Science, Tianjin University, 300350 Tianjin, China\\
2) Institute for Theoretical Physics, University of Amsterdam, 1090 GL Amsterdam, Netherlands}

\date{}

\abstract{
We study gauge theories between two parallel boundaries with non-trivial boundary conditions, which serve as a toy model for black hole background with two boundaries near the horizon and infinite, aiming for a better understanding of the Bekenstein-Hawking entropy. 
The new set of boundary conditions allows boundary modes and non-local modes that interplay between the two boundaries.
Those boundary modes and Wilson lines stretched between the two boundaries are carefully analyzed and are confirmed as physical variables in the phase space. 
Along with bulk fluctuation modes and topological modes, the partition function and entropy of all physical modes are evaluated via Euclidean path integral.
It is shown that there are transitions between the dominance of different modes as we vary the temperature.
The boundary fluctuation modes whose entropy is proportional to the volume dominate at high temperatures, and the boundary-area scaled boundary modes and Wilson lines are the more important at low temperatures. At super-low temperatures, when all the fluctuation modes die off, we see the topological modes whose entropy is the logarithm of the length scales of the system.
The boundary modes and non-local modes should have their counterparts in a black hole system with similar boundary conditions, which might provide important hints for black hole physics.
}


\maketitle


\section{Introduction}
\label{intro}

Gauge theories with non-trivial boundary conditions are important aspects of theoretical physics and can be used to understand lots of interesting physical phenomena \cite{Regge:1974zd, Wadia:1976fa, Gervais:1978kn, Barnich:2019qex, Barnich:2018zdg}. 
The would-be gauge degrees of freedom, which are no longer pure gauge, can become physical modes due to boundary conditions, which were suggested to explain the micro-states of black holes \cite{Carlip:1998wz, Donnelly:2014fua, Donnelly:2015hxa, Donnelly:2016auv, Barnich:2018zdg, Barnich:2019qex, Blommaert:2018oue, Blommaert:2018rsf}.
Moreover, most theoretical physicists tend to believe that the boundary degrees of freedom 
are vital for a better comprehension of the quantum effects of gravity, for example in the path integral formulation \cite{Gibbons:1976ue} and in AdS/CFT \cite{Stanford:2017thb, Mertens:2022ujr}. 

As suggested in \cite{Carlip:1998wz, Donnelly:2014fua, Donnelly:2015hxa, Donnelly:2016auv, Barnich:2018zdg, Barnich:2019qex, Blommaert:2018oue, Blommaert:2018rsf}, it is important to carefully study the boundary would-be gauge degrees of freedom. 
However, simply saying the boundary pure gauge configurations $\lambda_{\text{bdy}}$ are physical due to the boundary conditions and counting the corresponding entropy of those modes does not bring us anything. A relatively proper way to deal with those modes is to introduce a boundary current $J^{\mu}$ coupled with the boundary gauge fields in the action \cite{Cheng:2020vzw}. Functional integration of the current in the path integral naturally introduces an effective action for the boundary would-be gauge modes, which is more or less proportional to $(\pd_\mu \lambda_{\text{bdy}})^2$. Note that the above procedure is not gauge invariant and the non-gauge invariance of the key ingredient of the story.
The drawbacks of the above procedure are that boundary conditions are less transparent and the introduction of the boundary current seems to be artificial.

We are aiming to better understand the physics related to would-be gauge degrees of freedom from a set of nontrivial boundary conditions, where boundary modes and Wilson lines stretched between the two boundaries are allowed.
More specifically, we study gauge fields living between two parallel plates with non-trivial boundary conditions, shown in \eqref{second}, and carefully separate different parts of contributions in the presence of those boundaries.
The boundary condition we are interested in is the one where we allow residual degrees of freedom for the component of $A_\mu$ that is perpendicular to the boundaries to exist. 
The canonical formulation is carefully studied such that dynamical modes in the phase space (or Hilbert space) and the measure in the path integral is clear. The bulk fluctuation modes, boundary modes, bulk Wilson lines that stretched between two boundaries, and other topological modes should be considered as physical degrees of freedom in the current setup. The Wilson lines stretched between two boundaries are defined in equation \eqref{WL}, which captures the difference of boundary conditions on two boundaries.
Then, we evaluate the thermal partition function and entropy via the Euclidean path integral. 

As the temperature of the system varies from high temperatures to super-low temperatures, different modes dominate at different temperatures, and we can say that there are phase transitions between different modes.
The bulk fluctuation modes always dominate at high temperatures, whose entropy is proportional to the volume. The entropies of the boundary modes and Wilson lines are both proportional to the area of the boundaries, which should be useful in understanding the Bekenstein-Hawking entropy on black hole backgrounds. In super-low temperature limits, all the fluctuation modes play a less important role, and we can see interesting competition between the constant modes and topological modes. Those behaviors in the super-low temperature limits are supposed to provide some hints for the extremal limit on black hole backgrounds. A careful analysis of the black hole case is devoted to further studies. Note that we are considering a more general set of boundary conditions compared to previous studies \cite{Donnelly:2014fua, Donnelly:2015hxa, Donnelly:2016auv, Barnich:2018zdg, Blommaert:2018oue, Blommaert:2018rsf}. Rather than only considering boundary modes on specific boundaries, the presence of the boundary-stretched Wilson lines makes the structure of the phase space richer, which can also be important for black hole entropy and building connection with black hole soft hairs.

The relevant research can be an important aspect of understanding the microscopic interpretation of the Beken- stein-Hawking entropy. Like in the brick wall model built by 't Hooft \cite{tHooft:1984kcu}, there can be two boundaries: one near the horizon and the other at infinity. It was shown that black hole micro-states can be understood by studying the fluctuations (like a scalar field) on such backgrounds with two boundaries. The flat case we are going to study in this paper can be regarded as a toy model of the black hole case with gauge theories being considered, where effects due to the presence of two boundaries and non-local effects are the key ingredients of our model. 
We leave the study of the non-local effects on a black hole background for future research.
Moreover, the study of bulk gauge theory with two boundaries might be helpful to the recent progress on wormhole geometry \cite{Penington:2019kki, Almheiri:2019qdq} and the factorization puzzle \cite{Harlow:2019yfa, Blommaert2021, Saad:2021rcu, Saad:2021uzi, An:2023dmo, Cheng:2022nra}. We hope the non-local effects we study here can help us better understand the puzzles in AdS/CFT and quantum gravity.

The paper is organized as follows. In section \ref{bc}, we discuss the basic setup and interesting boundary conditions. The section \ref{canonical} is devoted to studying the canonical formulation of the theory, where we focus on the symplectic form and phase space. We also discuss the relationship between the canonical formulation and path integral, which leads to the Euclidean path integral part in section \ref{flatEPI}. We carefully analyze the contributions from different modes via path integral in section \ref{flatEPI}. In section \ref{temperature}, we demonstrate the transition between the dominance of different modes. Section \ref{con} is the conclusion section. 
The appendix \ref{boxtem} provides more details of the calculations in the main text.

This is the first paper in a series of papers that study the non-trivial boundary conditions of gauge theories. In the second paper of the series \cite{Cheng2302}, we study a similar set of boundary conditions on black hole background and are trying to understand the difference between the microstructures of finite-temperature and extremal black holes.

\section{Boundary conditions}
\label{bc}

\begin{figure}
\centering
\includegraphics[width=0.5\textwidth]{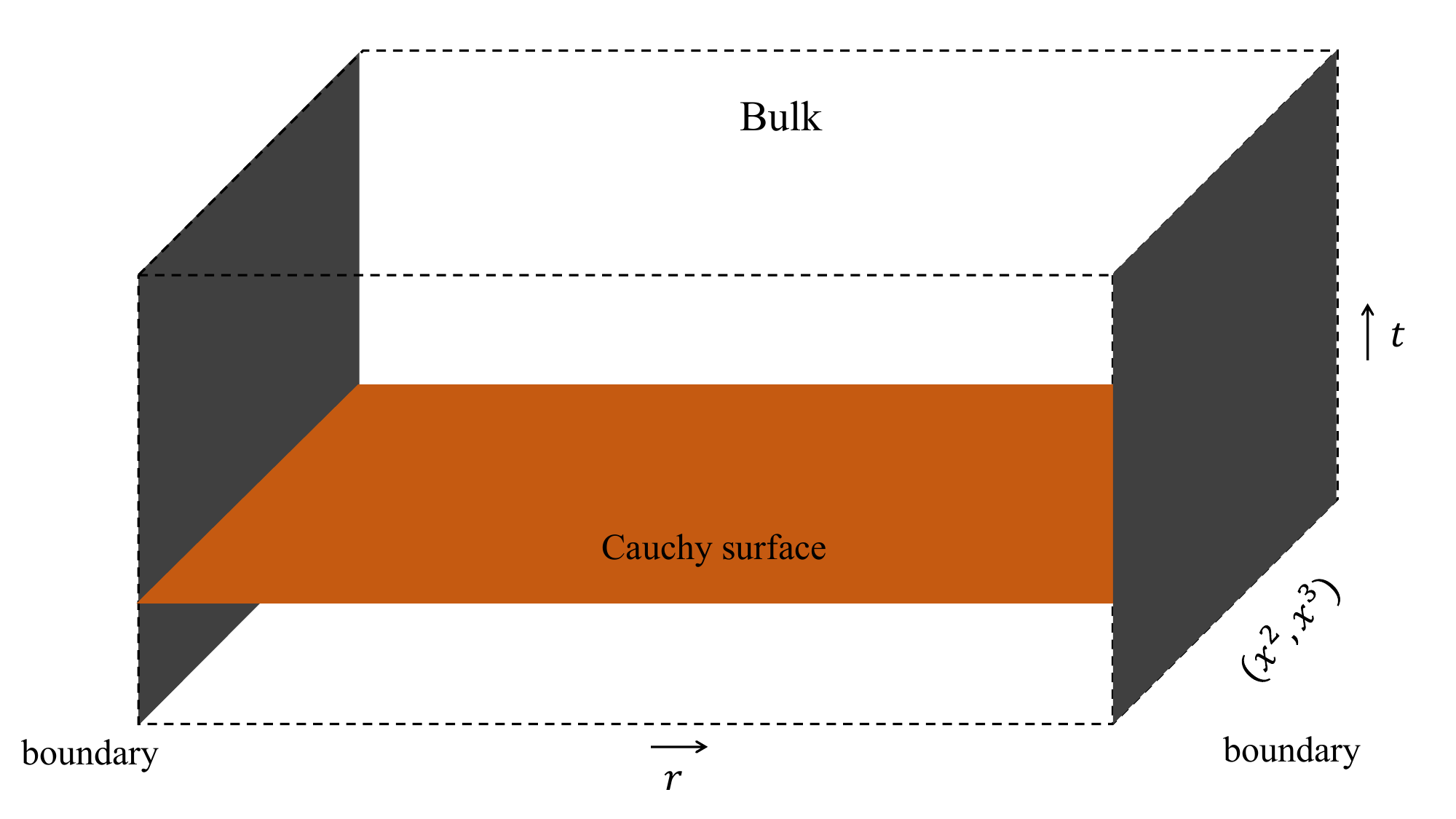}
\caption{U(1) gauge theory living between two parallel boundaries. The orange surface is a Cauchy surface with constant time.}\label{cauchy}
\end{figure}

This section studies the boundary conditions for a U(1) gauge theory living between flat parallel plates.
The situation we are mainly interested in is shown in figure \ref{cauchy}, where we have a Maxwell field theory living between two parallel boundaries on the left- and right-hand side. The original Maxwell theory has the action
\be
S=-\frac{1}{4e^2}\int_{\mM}d^4 x ~F^{\mu\nu}F_{\mu\nu}\,,\label{action1}
\ee
where $\mM$ is the 4-dimensional manifold, and $e^2$ is a dimensionless coupling constant. 
The 4-dimensional box has coordinate system $x^{\mu}=(x^a,r)=(t,x^2,x^3,r)$, where $r$ is the radius direction, and $x^a$ are the directions along the boundaries. 
To perform the finite temperature field theory calculations, we Wick rotate the time direction $t\to -i\tau$ such that $\tau$ becomes the Euclidean time with a periodicity $\beta$, i.e. the inverse temperature of the spacetime.

If all the gauge fields die off near the boundary, this is just the blackbody radiation with two polarisation degrees of freedom after gauge fixing. However, interesting phenomena start to show up when we release the boundary conditions, meaning that there are extra boundary degrees of freedom allowed due to the nontrivial boundary conditions.
Let us suppose the boundaries shown in figure \ref{cauchy} are labeled by $r= r_\alpha$, with the left and right boundaries located at $r_{(l)}=0$ or $r_{(r)}=L$.
$L$ is the distance between the two plates.
Then, in order to have a well-defined Hilbert space and variation principle, there would be some constraints for the boundary conditions. 
To see those constraints, let us first look at the variation of the action (\ref{action1}), which can be written as
\be
\delta S=\frac{1}{e^2}\int_{\mM}d\tau d^3x ~\pd_\mu F^{\mu\nu}\delta A_{\nu}-\frac{1}{e^2}\int_{\pd\mM}d\tau d^2x~n_{\mu}F^{\mu\nu}\delta A_{\nu}\,.\label{varS}
\ee
The on-shell variation of the action can be written as
\be
\delta S=-\frac{1}{e^2}\int_{\pd\mM}d\tau d^2x~n_\mu F^{\mu\nu}\delta A_{\nu}\,.\label{onshell0}
\ee
For the boundaries with normal vector $n^\mu\pd_\mu=\pd_r$, as shown in figure \ref{cauchy}, the variation can be written as 
\be
\delta S=-\frac{1}{e^2}\int_{\pd\mM}d\tau d^2x~ F^{ra}\delta A_{a}\,.\label{onshell}
\ee
To have a well-defined variation principle without adding any Gibbons-Hawking-like term, we have two obvious choices: 
\be
F^{ra}\big{|}_{\pd\mM}=0\,,~~~\text{or}~~~~\delta A_a\big{|}_{\pd\mM}=0\,.\label{aa}
\ee

We will detailedly discuss \eqref{aa} later. For now, let us look at the other boundary components, i.e. 
\be
n_{\mu} {}^*F^{\mu\nu}\big{|}_{\pd\mM}\,,~~~~~~\delta A_r\big{|}_{\pd\mM}\,.\label{bb}
\ee
Note that the variation principle has no constraint on the components shown in \eqref{bb}, and thus can be arbitrary at first sight. 
The physically motivated metallic Casimir boundary condition \cite{Milton2001, Bordag2009, Jaffe2005, Chernodub:2017gwe, Chernodub:2017mhi, Chernodub:2018pmt, Chernodub:2022izt, Chernodub:2023dok}, that requires the magnetic field normal to the boundary $\textbf{B}_{\perp}\big{|}_{\pd\mM}$ and electric field tangential to the boundary $\textbf{E}_{\parallel}\big{|}_{\pd\mM}$ to vanish. This is equivalent to asking 
\be
n_{\mu} {}^*F^{\mu\nu}\big{|}_{\pd\mM}=0\,.\label{casimir}
\ee
Essentially, we are interested in the boundary conditions that respect the physically motivated boundary condition \eqref{casimir}, while allowing $\delta A_r\big{|}_{\pd\mM}$ to be arbitrary.

Now, let us discuss the boundary conditions \eqref{aa} needed for a well-defined variation principle:
\begin{itemize}
\item For the first choice, we can have Neumann-like boundary condition
  \be\label{Neumann}
  F^{ra}\big{|}_{\pd\mM}=0\,.
  \ee
  Combined with \eqref{casimir}, we have $F^{\mu\nu}\big{|}_{\pd\mM}=0$. The situation doesn't seem super interesting, because the only possible allowed boundary configurations are the pure gauge configurations.

\item For the second choice  
  \be
  \delta A_a\big{|}_{\pd\mM}=0,
  \ee
  $F^{ra}\big{|}_{\pd\mM}$ can take arbitrary values. There can be interesting physics when we allow 
  \be
   \delta A_r\big{|}_{\pd\mM}\neq 0,. 
  \ee
 This is the situation we are going to investigate in the current paper.
\end{itemize}

So the boundary conditions we are mainly interested in are the ones that consistent with 
  \be\label{second}
  \delta A_a\big{|}_{\pd\mM}=0,~~~~~~~\delta A_r\big{|}_{\pd\mM}=f(x^a)\,.
  \ee
where $f(x^a)$ can have local dependence of $x^a$. 
  $A_a\big{|}_{\pd\mM}$ are fixed configurations on boundaries, which don't need to be summed over in path integral. 
Meanwhile, there is no constraint on $A_r$ at the boundary, and correspondingly, $F^{ra}\big{|}_{\pd\mM}$ can be arbitrary. The Hilbert space is well-defined with the fixed boundary configurations $A_a\big{|}_{\pd\mM}$, and we need to sum over different boundary configurations of $A_r$ in path integral.

Let us see what are the configurations that respect \eqref{second}. For the $A_a$ components, the boundary configurations are fixed and we can let those fields be zero at the boundaries
\bea\label{Aabc}
A_a\big{|}_{r=r_\alpha} = 0\,.
\eea
Note that the above setting also respects the Casimir boundary condition \eqref{casimir}.
$A_r\big{|}_{\pd\mM}$ can take different configurations at the left and right boundaries. 
Besides, the boundary configurations can fluctuate and have arbitrary $x^a$ dependence. So we need to separate the bulk and boundary configurations carefully. 
$A_r$ can be separated as follows
\be
A_r(x^\mu)=\hat{A}_r(x^\mu) +\frac{\phi(x^a)}{L}\,,
\ee
where $\phi(x^a)$ are the configurations that makes $\hat{A}_r\big{|}_{r=0}=0$, i.e. $A_r(r,x^a)\Big{|}_{r=0}=\phi(x^a)/L$.
$L$ is added mainly for the dimension-counting reason. To write the fields in a more concordant way, we can also use $\hat{A}_{a}=A_a$ to denote the bulk configurations that vanish on boundaries.
Note that with the above separation, $\hat{A}_r$ does not equal zero at the boundary $r=L$, and we can also further decompose $\hat{A}_r$ into two parts.
The part that satisfies $\int_0^L dr~ \hat{A}_r=0$ will not be our main concern here, and the part that satisfies  $\int_0^L dr~ \hat{A}_r\neq 0$ will be an important ingredient in our later calculations.
One can also decompose $\hat{A}_r$ into the part that vanishes on both sides and the part that captures the difference between the two boundaries. Integrating the part that vanishes on both boundaries from one boundary to the other gives out zero, and $\int_0^L dr~ \hat{A}_r$ more or less captures the difference of $A_r$ between the two boundaries.
The reason why we don't choose this decomposition is that from the bulk point of view, we need bulk on-shell configurations corresponding to those boundary degrees of freedom. The corresponding bulk configurations need to be on-shell such that we have a good separation of different modes in the action. 
What's more, as we will see in the next section, $\int_0^L dr~ \hat{A}_r$ has a good physical interpretation.

Note that as already motivated in the introduction, we allow non-gauge invariance in the system to capture the possible would-be gauge modes that are wildly interested in gravitational systems. It is okay to add the would-be gauge modes that respect the boundary condition into the system as far as there are confirmed as physical through canonical analysis. So, before we actually integrate all the degrees of freedom in the Euclidean path integral, let us first analyze the theory's canonical phase space and see which degrees of freedom are physical. Only the physical degrees of freedom are supposed to be integrated over in the path integral. This is what we are going to do in the next section.

\section{Canonical formulation}
\label{canonical}

\begin{figure}
\centering
\includegraphics[width=0.48\textwidth]{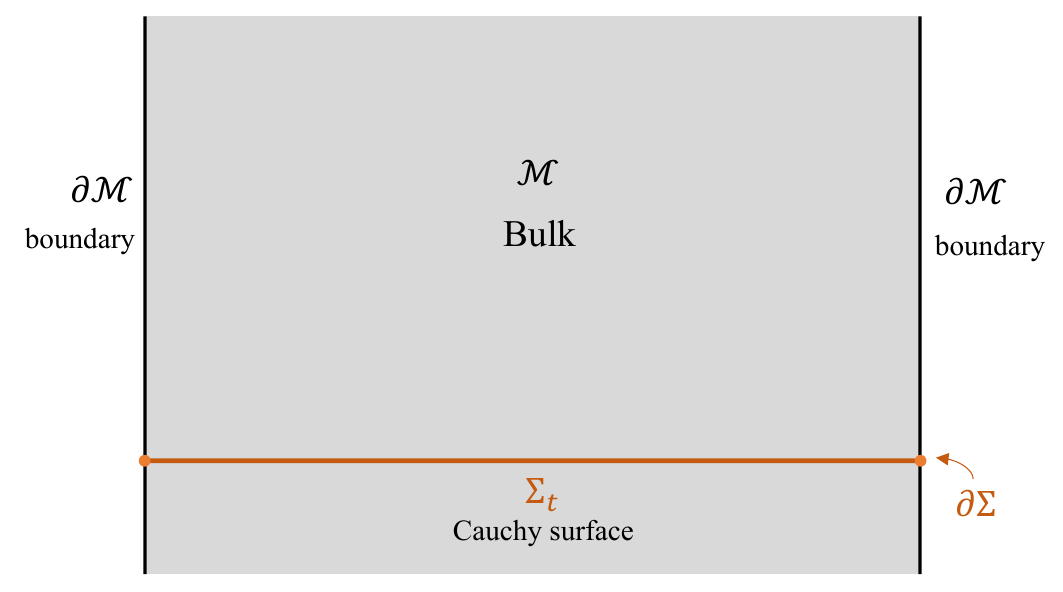}
\caption{The bulk manifold $\mM$, the boundaries $\pd\mM$, Cauchy surface $\Sigma_t$, and the boundaries of the Cauchy surface $\pd\Sigma$ are all illustrated here.}\label{notation}
\end{figure}

We are going to work out the phase space and the equipped symplectic form in canonical formulation for the theory with non-trivial boundary conditions \eqref{second}.
The canonical formulation is always related to a Cauchy surface where the Hilbert space is defined. For the flat parallel plates case, a Cauchy surface is shown in Fig. \ref{notation}. The canonical formulation can help us to better understand what are the physical degrees of freedom in the phase space. 
The phase space can be represented by $\Gamma$, which is an even-dimensional manifold with coordinates $x^I=\{q^i,p_j\}$, where $q^i$ and $p_j$ are the canonical coordinates and momenta.
For field theories, we have infinite-dimensional phase spaces $\Gamma$.
After canonical quantization, the phase space can be turned into the Hilbert space of the theory.

 For U(1) gauge theory with trivial boundary conditions, we can decompose the gauge fields into temporal and spatial directions and rewrite the gauge fields $A_{\mu}$ as
 \be
 A_{\mu}=(-V, A_i),~~~~A^{\mu}=(V,A^i)\,,
 \ee
we have the Lagrangian density written in terms of $V$ and $A_i$ as
\be
\mL=\frac{1}{2e^2}(\dot A^i+\pd^i V)(\dot A_i+\pd_i V)-\frac{1}{2e^2}F^{ij}\pd_i A_j\,.
\ee
The corresponding conjugate momenta of fields $V$ and $A_i$ can be written as
\be
\Pi_V=\frac{\pd \mL}{\pd \dot V}\,~~~~\Pi^i=\frac{\pd \mL}{\pd \dot A_i}=\frac{1}{e^2}(\dot A^i+\pd^i V)\,,\label{momenta}
\ee
where we have denoted $\Pi_V$ as the momentum for $V$ and $\Pi^i$ as the momenta for $A_i$.
So, for trivial boundary conditions, the phase space we start with is 
\be\label{gamma0}
\Gamma_0=\{V, A_i, \Pi_V, \Pi^i\}\,.
\ee
Gauge fixing conditions help us further get rid of the unphysical degrees of freedom in the phase space and we end up with the standard two polarizations of photon in Maxwell's theory. 
Note that it is natural to use temporal gauge $A_t=0$ in the canonical formulation.

However, there can be boundary subtleties in the phase space when we have nontrivial boundary conditions, like \eqref{second}. 
What are those boundary subtleties?
This can be answered by turning to the symplectic form of the theory and working out the Poisson bracket between the fields to make those boundary subtleties more explicit. 
Moreover, an explicit phase space and symplectic form can help us figure out the canonical variables and momentum, which should be integrated over in the path integral.
The phase space is equipped with a closed, non-degenerate symplectic two-form $\Omega$, which is defined as
\be
\Omega =\frac{1}{2}\Omega_{IJ}\dd x^I\wedge \dd x^J\,.
\ee
$\Omega_{IJ}$ is invertible, and the inverse $\Omega^{IJ}$ is defined by $\Omega^{IK}\Omega_{KJ}=\delta^I_J$. Now equipped with the symplectic form, the classical Poisson bracket between functionals $F$ and $G$ can be defined as
\be
\{F,G\}=\Omega^{IJ}\frac{\delta F}{\delta x^I}\frac{\delta G}{\delta x^J}\,.
\ee
Quantum commutators can be obtained by the canonical quantization procedure.

The symplectic form of a field theory can be directly worked out on a chosen Cauchy surface \cite{Wald:1999wa}.
Let us consider a field theory with a Lagrange density $\mL[\Psi]$, where $\Psi$ denotes an arbitrary collection of fields. Taking the variation of $\mL$, we have
\be\label{var}
\delta \mL=E \cdot \delta \Psi+\dd \Theta\,.
\ee
The equation of motion $E=0$ kills the first term in \eqref{var}. The (pre)-symplectic potential $\Theta[\Psi,\delta \Psi]$ is a $D-1$ form and can be integrated over the chosen $(D-1)$-dimensional Cauchy surface. The symplectic current $\omega$ can be defined as
\be
\omega[\Psi,\delta_1 \Psi, \delta_2 \Psi]=\delta_1 \Theta[\Psi,\delta_2 \Psi]-\delta_2 \Theta[\Psi,\delta_1 \Psi]\,,
\ee
where $\delta_1$ and $\delta_2$ can be regarded as variations with respect to two different transformations. Integrating the symplectic current $\omega$ over the Cauchy surface $\Sigma$, we finally get the symplectic form $\Omega$ written as
\be
\Omega[\Psi,\delta_1 \Psi, \delta_2 \Psi]=\int_{\Sigma}\omega[\Psi,\delta_1 \Psi, \delta_2 \Psi]\,.
\ee
Note that the choice of the $(D-1)$-form $\omega[\Psi,\delta_1 \Psi, \delta_2 \Psi]$ also depends on the Cauchy surface. Specifying to the temporal surface $\Sigma_t$ with normal vector $n^t$, we can work out each component of $\omega$.


Now, let us be specific to the U(1) gauge theory with non-trivial boundary conditions at hand.
The variation of the action is similar to \eqref{varS}. The difference is $\Theta$, as well as the symplectic form, is defined on Cauchy surface $\Sigma$, as shown in Fig. \ref{notation}.
Thus, the symplectic form $\Omega_{\Sigma}$ can be worked out as
\bea
\Omega_{\Sigma} = -\frac{1}{e^2}\int_{\Sigma}d^3x ~ n^{\mu}\delta F_{\mu\nu}\wedge \delta A^\nu\,.
\eea
Specifying on the chosen Cauchy surface $\Sigma_t$ with normal vector $n^\mu\pd_\mu=\pd_t$, the above expression can be written in components as
\be
\Omega_{\Sigma_t}=-\frac{1}{e^2}\int_{\Sigma_t}d^3x~\delta F_{ti}\wedge\delta A^i\,.\label{sym}
\ee

The symplectic form is essential in defining the Hamiltonian dynamics; only the fields equipped with a nontrivial symplectic form with their conjugate momentum can be regarded as dynamical variables in the phase space.
We have decomposed the gauge fields into several parts as we discussed at the end of the section \ref{bc}. 
We are going to see what are the symplectic partners of all the configurations in the symplectic form.

Now one can put the variation of fields into the symplectic form (\ref{sym}), and rewrite it as
\bea
\Omega_{\Sigma_t} &=& -\frac{1}{e^2}\int_{\Sigma_t}d^3x ~\Big [\delta( \hat F_{tr}+\frac{\dot \phi}{L})\wedge \delta (\hat{A}^r+\frac{\phi}{L})\nonumber \\ 
&~&~~~~~~~~~~~~~~~~~~+\delta \hat F_{t2}\wedge \delta \hat{A}^2+\delta \hat F_{t3}\wedge \delta \hat{A}^3 \Big]\,.
\eea
We can separate the $\hat A_i$ part with other parts, and the symplectic form reads as
\bea\label{sym1}
\Omega_{\Sigma_t} &=& -\frac{1}{e^2}\int_{\Sigma_t}d^3x ~\delta \hat F_{ti}\wedge \delta \hat{A}^i
\,\\ &~&
 -\frac{1}{e^2}\int_{\Sigma_t}d^2x dr~\Big[\frac{\delta\dot \phi}{L}\wedge \delta \hat{A}^r+\frac{\delta\dot \phi}{L}\wedge\frac{ \delta \phi}{L}+\delta \hat F_{tr}\wedge \frac{\delta \phi}{L}\Big]\,.\nn
\eea
The first term in (\ref{sym1}) gives us the usual Poisson bracket of Maxwell's theory.
Integrating over $r$ in the second term gives out
\bea\label{sbdy}
\Omega_{\text{bdy}} &=&-\frac{1}{L\cdot e^2}\int d^2x ~\Big[\delta \dot\phi \wedge \delta (\int_0^L dr \hat A^r)+\delta \dot\phi \wedge \delta \phi\nn\\
&~&~~~~~~~~~~~~~~~~~~~~~~+\delta (\int_0^L dr \dot{\hat A}_r)\wedge \delta \phi
\Big]\,,
\eea
where we have used the boundary condition $\delta \hat A_t\big{|}_{\pd\Sigma}=0$. $\Omega_{\text{bdy}}$ can be regarded as the symplectic form of the dynamical variables due to the presence of boundary condition \eqref{second}.

The above symplectic form tells us what are the extra physical degrees of freedom in the system, besides phase space \eqref{gamma0}.
The non-local part in \eqref{sbdy} can be denoted as a separate variable. 
Inspired by that, we define the quantity $W$ as
\be\label{WL}
W=i\int_0^L dr~ \hat A_r\,,
\ee
which will be called \textit{Wilson lines}\footnote{
The actual Wilson lines stretched between the two boundaries can be written as
\be
\mathcal {W}\propto \mathcal {P}\exp \left[i(\int_0^L dr~ \hat A_r)+i\phi \right]\,,
\ee
with $\mathcal {P}$ denoting the path ordering. $\phi$ is the boundary configuration of $A_r$. The definition of the Wilson line can be exactly matched if we choose a different separation of degrees of freedom as discussed at the end of section \ref{bc}. 
}.
Note that field $W$ captures the difference of $A_r$ on the two boundaries. From now on, we will extract $W$ modes out of $\hat A_r$ such that $\hat A_r$ equals zero at both boundaries.

The boundary symplectic form, defined on a codimension-2 surface, can be written as
\be
\Omega_{\text{bdy}}=
-\frac{1}{e^2 L}\int d^2x ~\left[-i\delta \dot\phi \wedge \delta W -i\delta \dot W \wedge \delta \phi +\delta \dot\phi \wedge \delta \phi \right]\,.
\ee
The cross-term between $W$ and $\phi$ can be canceled by refining the fields. For example, shifting $\phi \to \phi +iW$, the above symplectic form can be written as
\be
\Omega_{\text{bdy}}=-\frac{1}{e^2 L}\int d^2x ~\left[\delta \dot\phi \wedge \delta \phi +\delta \dot W \wedge \delta W \right]\,.
\ee
With the above redefinition of fields, the overall symplectic form \eqref{sym1} can be expressed as
\bea
\Omega_{\Sigma_t} &=&-\frac{1}{e^2}\int_{\Sigma_t}d^3x~ \delta \hat{F}_{ti}\wedge \delta \hat A^i  \nn\\
&~&-\frac{1}{e^2 L}\int d^2x ~\left[\delta \dot\phi \wedge \delta \phi +\delta \dot W \wedge \delta W \right]\, \nn\\
&=&-\frac{1}{e^2}\int_{\Sigma_t}d^3x~ \delta \hat\Pi_i \wedge \delta \hat A^i  \nn\\
&~&-\frac{1}{e^2 L}\int d^2x ~\left[\delta \Pi_W \wedge \delta W+ \delta \Pi_\phi \wedge \delta \phi\right]\,,\label{symplectic}
\eea
where $\hat\Pi^i$ denotes the conjugate momentum of $\hat A_i$.
Note that, beside the conjugate momentum defined in \eqref{momenta}, we further define the conjugate momentum of $W$ and $\phi$ as
\be
\Pi_W=\dot W,~~~~\Pi_\phi=\dot\phi\,.
\ee
The Poisson brackets can be derived as
\begin{align}
	\label{Poisson1}
&\frac{1}{e^2}[~\hat\Pi^i(r,x^2,x^3),\hat A_j(r',x'^2,x'^3)~] =i\delta^i_j ~\delta(r-r')~\delta^2(x-x')\,,\\
&\frac{1}{e^2 L}[~\Pi_W(x^2,x^3),W(x'^2,x'^3)~] = i\delta^2(x-x')\,,\label{Poisson2}\\
&\frac{1}{e^2 L}[~\Pi_\phi(x^2,x^3),\phi(x'^2,x'^3)~] = i\delta^2(x-x')\,.\label{Poisson3}
\end{align}

As discussed at the beginning of this section, we need to add the degrees of freedom related to the boundary subtleties back to the phase space and perform path integration over those configurations in the path integral. Those boundary subtleties are the degrees of freedom related to the boundary configurations of $A_r$.
By the detailed symplectic form analysis in this subsection, we have found the zero modes $\phi(x^a)$ that take zero value for the longitudinal momentum along $r$ direction and Wilson lines stretched between the two boundaries $W$ have non-trivial symplectic partners and Poisson brackets. Those are the degrees of freedom needed to be added back. So the actual phase space should be
\be
\Gamma=\left\{~\hat\Pi^i, \hat A_i, \Pi_\phi, \phi, \Pi_W, W~ \right\}\,.
\ee
Further gauge fixing conditions would help us to get rid of the bulk gauge redundancy of $\{ \hat\Pi^i, \hat A_i\}$ such that we are only left with two bulk polarizations. 
This isn't our main concern in this paper.
The new ingredients are $\phi$ and $W$, where $\phi$ is the zero modes along the $r$ direction of $A_r$, and $W$ is the Wilson lines stretched between those two boundaries.

As discussed at the beginning of the paper, the modes that appear in the canonical formula should be included in the path integral method of calculating the partition function.
It can easily be seen that the partition function $Z=$tr $e^{-\beta H}$ with the Hamiltonian $H$, can be written as a path integral over $\tau \in [0,\beta]$, with the compact Euclidean time $\tau=it$ (see e.g. \cite{Kapusta2011} for more details).
With the relation between Hamiltonian formula tr $e^{-\beta H}$ and Euclidean path integral $\int \mD \vpp~ e^{-S_E}$ being clear, we can apply all the canonical analysis to the path integral formalism. 

In the next section, we are going to include all the physical degrees of freedom into the Euclidean path integral to evaluate their contribution to entropy.
Note that the canonical analysis gives us some hints about what degrees of freedom are physical. Although we can use the canonical results as input and use the relation between the canonical formula and the Euclidean path integral to work out the partition function, we will study the path integral carefully. We may use different gauge fixing conditions if they are more convenient.

\section{Euclidean path integral}
\label{flatEPI}

After the canonical analysis in the previous section, it is clear that the dynamical modes are the bulk fluctuation modes $\hat A_\mu$, zero modes along $r$ direction $\phi$, and the Wilson lines stretched between the two boundaries $W$. Those fields are the ingredients that need to be included in the Euclidean path integral.
Note that there should only be two bulk physical polarizations for the fields $\hat A_\mu$ after gauge fixing. Specially caution is needed when handling gauge fixing, and we will deal with bulk gauge fixing conditions after the physics are clear to avoid gauging too much or too little.

As discussed in the previous subsection, the partition function can be written as a Euclidean path integral
\be
Z=\int \mD A_\mu ~e^{-S_E}\,,\label{Zoriginal}
\ee
with the Euclidean action $S_E$ written as
\be\label{SE0}
S_E=\frac{1}{4e^2}\int_\mM d\tau d^3x~ F^{\mu\nu}F_{\mu\nu}\,.
\ee
The above is the original formula, and we are going to do some massage according to the hints from the canonical analysis.
We can separate the $x^a$ directions with $r$ in the action
\bea
S_E = \frac{1}{4e^2}\int_\mM d\tau d^3x~ F^{ab}F_{ab}+\frac{1}{2e^2}\int_\mM d\tau d^3x~ F^{ra}F_{ra}\,.
\eea
Again, we are going to separate the gauge fields into different parts
\bea
A_a = \hat A_a\,,~~~~~~
A_r = \hat A_r+\frac{\phi(x^a)}{L}\,.
\eea
The Euclidean action can be written in terms of those modes as
\bea
S_E 
&=&
\frac{1}{4e^2}\int_\mM d\tau d^3x~ \hat F^{\mu\nu}\hat F_{\mu\nu} \nn\\
&~&+\frac{1}{2e^2}\int_\mM d\tau d^3x~ \left[ -\frac{2}{L}~ \hat F^{ra} \pd_a \phi+\frac{\pd^a \phi\pd_a\phi}{L^2} \right]\,.
\eea
The second part of the above expression contains the interesting ingredients of our story and can be denoted as
\be
S_{\text{bdy}}=\frac{1}{2e^2}\int_\mM d\tau d^2x dr~ \left[ \frac{\pd^a \phi\pd_a\phi}{L^2}+\frac{2}{L}~ \hat F^{ar} \pd_a \phi\right]\,.
\ee
$\phi(x^a)$ is not a function of radius direction $r$, so the integral over $r$ can just pass through  $\phi$, and gives out an extra $L$ in the first term. Noticing that $\hat A_a$ equal zero at the boundaries $\pd \mM$, integrating over $r$ in the above effective action gives out
\bea
S_{\text{bdy}} 
&=& \frac{1}{2e^2 L}\int d\tau d^2x~ \Big[ {\pd^a \phi\pd_a\phi} -2i~ \pd^a (i\int dr \hat A^r)  \pd_a \phi \Big ]\,.\nn\\
\eea
Denoting $W(x^a) = i\int dr \hat A_r$, the original action can be rewritten as
\bea\label{effectS}
S_E &=& \frac{1}{4e^2}\int_\mM d\tau d^3x~ \hat F^{\mu\nu}\hat F_{\mu\nu}\nn\\
&~& +\frac{1}{2e^2 L}\int d\tau d^2x~ \left[ {\pd^a \phi\pd_a\phi} -2i~ \pd^a W \pd_a \phi \right]\,.
\eea

We have several remarks regarding the Euclidean path integral and the effective action \eqref{effectS}:
\begin{itemize}
  \item The original path integral \eqref{Zoriginal} is a path integral over $A_\mu$, while we have different variables in action \eqref{effectS}. There is no difference between $\hat{A}_a$ and $A_a$. 
The integral over $A_r$ component can be divided into several different pieces. 
The zero modes $\phi$ and the Wilson lines $W$ are the parts capturing the boundary configurations of $A_r$. The bulk modes $\hat A_r$ that satisfy the conditions $\hat A_r\big{|}_{r=0}=0$ and $\int_0^L dr \hat A_r=0$ should be regarded as the bulk contribution. We can always gauge fix $\hat A_r$ to zero, which doesn't kill any important physics.

  \item One of the main purposes of the canonical analysis in the previous section is to make clear the measure of different modes in the path integral. From the symplectic form \eqref{symplectic} and Poisson brackets (\ref{Poisson1}-\ref{Poisson3}), the measure can be easily determined.
  
  \item Putting the effective action \eqref{effectS} in the path integral \eqref{Zoriginal}, we can first work out the Gaussian integral over $\phi$, which gives out $\det(\pd^2)^{-1/2}$ in the partition function.
The above procedure also gives out an effective action for $W$, which is the action for a 3-dimensional massless scalar field.
The determinant $\det(\pd^2)^{-1/2}$ getting from integrating $\phi$ can be rewritten as path integral. 
With all the above arguments, the effective action can be expressed as
\bea\label{ES}
S_E &=&\frac{1}{4e^2}\int_\mM d\tau d^3x~ \hat F^{\mu\nu}\hat F_{\mu\nu} \nn\\
&~& +\frac{1}{2e^2 L}\int d\tau d^2x~ \left[ {\pd^a \phi\pd_a\phi} + \pd^a W \pd_a W \right]\,.
\eea
One can also check the path integrals with actions \eqref{effectS} and \eqref{ES} give out the same result. So we will use \eqref{ES} as the effective action in evaluating the partition function later on.
As a direct analogy of the canonical analysis, the cross term between fields $W$ and $\phi$ in \eqref{effectS} can also be canceled by shifting $\phi \to \phi+iW$, which would give out the same result as \eqref{ES}.

  \item We have been super-careful about the gauge fixing such that we didn't gauge fix any interesting physics. For example, we can gauge fix part of the bulk fields $\hat A_{\mu}$ later, but we always need to make sure $\phi$ and $W$ are not gauged away.
There might be other interesting modes that are needed to be added back.
Let us suppose we are dealing with a compact U(1) gauge theory. In the Euclidean background, the map between the background time circle $\tau\sim \tau+\beta$ and the compact gauge parameter allows us to include some topological modes for component $A_\tau$. The fundamental group of $S_1$ is $\textbf{Z}$. So the modes can be expressed as
\be
A_{\tau} \ni \frac{2\pi n}{\beta}\,,~~~~~n\in \textbf{Z}\,.\label{largeGT}
\ee
Those modes correspond to the large gauge transformation and might be physical. 
Those modes respect the boundary condition \eqref{second}, but not \eqref{Aabc}. So, we do not include the modes \eqref{largeGT} in the current calculation because of the more strict boundary condition \eqref{Aabc}. 
\end{itemize}

Now, let us evaluate the Euclidean path integral. 
For the first term corresponding to a Maxwell theory with vanishing boundary conditions, we can denote the partition function as $Z_{\hat A}$.
The field $\phi$ and $W$ are 3-dimensional scalar fields living on a surface with coordinate $x^a$, which will be regarded as the boundary contribution.
We can separate the path integral as bulk and boundary parts
\bea
Z
&=& Z_{\hat A} \times   \int\mD \phi~ \mD W~ \nn\\ &\times&\exp\left[-\frac{1}{2e^2 L}\int d\tau d^2x (\pd^a \phi \pd_a\phi+ \pd^a W \pd_a W)\right]\,.
\eea
The main task left is to evaluate the bulk and boundary partition functions. We are going to discuss those different modes for the remainder of this section, and evaluate the partition function and demonstrate the possible phase transitions in the next section. 

\subsection{Bulk fluctuation modes}

First of all, let us evaluate the partition function for bulk fluctuation modes $Z_{\hat A}$. We will use the Faddeev-Popov method \cite{Faddeev:1967fc} to evaluate the partition function, by inserting the following identity
\be
1=\int \mD \lambda \det \left(\frac{\pd G}{\pd \lambda}\right)\delta(G-0)\,,\label{gaugefix}
\ee
with gauge fixing condition $G=\pd_\mu \hat A^{\mu}-c(x)$. Following the standard gauge fixing procedure, in Feynman gauge, we eventually get
\bea
Z_{\hat A} &=& \int \mD \hat A_{\mu}\mD C\mD \bar C~e^{-\frac{1}{2e^2}\int_\mM d\tau d^3x~[ \hat A^{\mu}(\pd^2 )\hat A_{\mu}+ \bar C (\pd^2)C]}\nn\\
&=& \det(\pd^2)^{-1}\,,\label{ZAfixed}
\eea
where $C$ and $\bar C$ are ghost fields. After gauge fixing, the final result is the partition function for two bosonic polarizations 
\be
Z_{\hat A} =\det(\pd^2)^{-1/2}\times\det(\pd^2)^{-1/2}
\ee
If we define the energy and momenta of the gauge fields as $(\omega,p_r,p_2,p_3)$, the logarithm of $Z_{\hat A}$ can be calculated by working out the determinantal operator
\bea
\ln Z_{\hat A}
= -\sum_{\omega}\sum_{p_r,p_2,p_3} \ln  \left[\beta^2(\omega^2+p_r^2+p_2^2+p_3^2)\right]\,.\label{lnZa}
\eea
One can further evaluated the partition function by taking different limits of the length scales in the theory. We are going to evaluate the logarithm of the partition function in section \ref{temperature} when we discuss different temperature limits.

Note that we can also use different gauge fixing conditions, like the axial gauge or temporal gauge. The gauge fixing condition does not make much difference for the fluctuation modes as far as we keep two physical polarizations in the final result.

\subsection{Fluctuation modes of $\phi$ and $W$}

Let us evaluate the partition function of fields $\phi$ and $W$ here.
The action for $\phi$ and $W$ can be written as
\be
S_{\phi,W}=\frac{1}{2e^2 L}\int d\tau d^2x ~(\pd_a \phi\pd^a \phi+\pd_a W\pd^a W)\,,
\ee
which is the action for two massless scalar fields living on the boundary.
Denoting the area of the boundary as ``$\A$", the fluctuation modes of field $\phi$ and $W$ can be expanded as
\bea
\phi(x^a) &=& \sqrt{\frac{2e^2 \beta L}{\A}} \sum_{\omega,p_2,p_3}\tilde{\phi}(\omega,p_2,p_3)e^{i(\omega \tau+p_2x^2+p_3x^3 )}\,,\nn
\\
W(x^a) &=& \sqrt{\frac{2e^2 \beta L} {\A}}\sum_{\omega,p_2,p_3}\tilde W(\omega,p_2,p_3)e^{i(\omega \tau+p_2x^2+p_3x^3 )}\,.\nn
\eea
The coefficient is chosen such that $\tilde \phi$s and $\tilde W$s are dimensionless and thus the integrals over $d \tilde \phi$ in the path integral give out dimensionless quantities.
With this mode expansion, the corresponding partition function can be expressed as
\be
Z_F=\prod_{\omega,p_2,p_3}[\beta^2(\omega^2+p_2^2+p_3^2)]^{-1}\,,
\ee
the logarithm of which is
\be
\ln Z_F=-\sum_{\omega,p_2,p_3}\ln[\beta^2(\omega^2+p_2^2+p_3^2)]\,.
\ee
$Z_F$ is the partition function for two 3-dimensional massless scalar fields. We can then calculate the free energy and entropy of those modes in different temperature limits, and compare it with the bulk fluctuation modes, which will be the task for the next section.

\subsection{Other interesting modes}

There are some other interesting topological modes of $W$.
The Wilson lines stretched between the two boundaries can be denoted as
\be
\mathcal{W}_\gamma=\mathcal{P}\exp[i\int_0^L dr \hat A_r]\,.
\ee
Because it's always inside of an exponential function, $\int_0^L dr \hat A_r$ is compact with periodicity $2\pi$.
The requirement that the Wilson lines are single-valued allows us to include the elements of fundamental group ${S}^1$.
In the Euclidean background, the background time circle $\tau\sim\tau+\beta$ allows the field $W$ to wind around the ${S}^1$ circle and have some winding modes ${2\pi n \tau }/{\beta}$.

Now the field $W$ has compact constant modes and novel winding modes which are interesting to deal with. The constant modes contribution of $W$ can always be written as
\be
Z_0=\int_0^{2\pi \sqrt{\frac{\A}{e^2L \beta }}}d\tilde W_0=2\pi \sqrt{\frac{1}{e^2 L}}\times\sqrt{\frac{\A}{\beta}}\,.
\ee
The winding mode contribution can be written as
\be
Z_w=\sum_n e^{-\frac{\A}{2e^2 L \beta}(2\pi n)^2}\,.
\ee
$Z_w$ equals 1 when the coefficient $\frac{\A}{2e^2 \beta L}$ is very large, since the mode with $n=0$ dominants. When the coefficient is very small, we can change the sum into a Gaussian integral. 
The overall partition function of $\phi$ and $W$ is the product of constant modes $Z_0$, winding modes $Z_w$, and fluctuation modes $Z_F$ discussed previously.

As a summary of the different modes and corresponding partition functions we have got, we can express the overall partition function as
\be\label{logZ}
\ln Z=\ln Z_{\hat A}+\ln Z_{F}+\ln Z_0+\ln Z_w\,.
\ee
We have two bulk polarizations in $Z_{\hat A}$, two collections of fluctuation modes from $W$ and $\phi$ in $\ln Z_{F}$, constant modes $Z_0$ and winding modes $Z_w$. The logarithm of the overall partition function can be expressed as
\bea
\ln Z &=& -2\times\frac{1}{2}\sum_{\omega,p_r,p_2,p_3}\ln[\beta^2(\omega^2+p_r^2+p_2^2+p_3^2)]\nn\\
&~&-2\times\frac{1}{2}\sum_{\omega,p_2,p_3}\ln[\beta^2(\omega^2+p_2^2+p_3^2)]\nn\\
&~&+\frac{1}{2}\ln[\frac{\A}{\beta L}]-\ln e+\ln \sum_n e^{-\frac{\A}{2e^2 L \beta}(2\pi n)^2}\,,\label{lnZ_phiN}
\eea
We will directly evaluate different parts of \eqref{logZ} in the next section.

\section{Transition between different phases}
\label{temperature}

In this section, we evaluate and compare the partition function shown in \eqref{logZ} in different temperature limits. The partition function contains contributions from the bulk fluctuation modes $\hat A_\mu$, the zero modes $\phi$, and the Wilson lines $W$. The detailed calculations of the partition function are included in Appendix \ref{boxtem}, such that we don't drown in the tasteless details. The main content of this section is devoted to the discussion of different behaviors and phase transitions.

There are three different dimensional length scales in the theory, the inverse temperature $\beta$, the distance between the two boundaries $L$, and the length scale of the boundary $\sqrt{\A}$. The coupling constant $e^2$ is dimensionless. We are going to compare the inverse temperature $\beta$ with other length scales in the theory and call them different temperature limits. In the different temperature limits, we can study different behaviors of the partition function.
We are mainly interested in the following three different temperature limits. 
\begin{itemize}
  \item The so-called \textit{high-temperature limit} is the limit when we have $\beta\ll L \ll\sqrt{\A}$. $\beta$ is the smallest length scale in the system. In this temperature limit, the bulk fluctuation modes $\hat A_\mu$ should be the most important contribution.
  \item The second temperature limit we are interested in is the \textit{low-temperature limit}, where we have $L\ll\beta\ll\sqrt{\A}$. The distance between the two boundaries is way smaller than the inverse temperature $\beta$, and all the high-frequency modes along the $r$ direction will be gapped.
The zero modes $\phi$ and the Wilson lines $W$ start to play the most important role in this limit.
  \item The last case is the \textit{super-low temperature limit}, where we have $L\ll\sqrt{\A}\ll\beta$. The temperature is super low, and all the fluctuation partition functions that proportional to the temperature are disappeared.  The logarithm contributions of fields $W$ shown in the previous section become the most important ones.
\end{itemize}

Let us discuss those three temperature limits separately. The qualitative behavior of the entropy is illustrated in figure \ref{boxphase}. The overall entropy is a summation of different contributions, while figure \ref{boxphase} illustrates the contributions from different modes.  The solid red curves show the dominant contributions. More details of the calculations can be found in Appendix \ref{boxtem}, the main results and behaviors are discussed below.

\begin{figure}
  \centering
  ~~\includegraphics[width=0.43\textwidth]{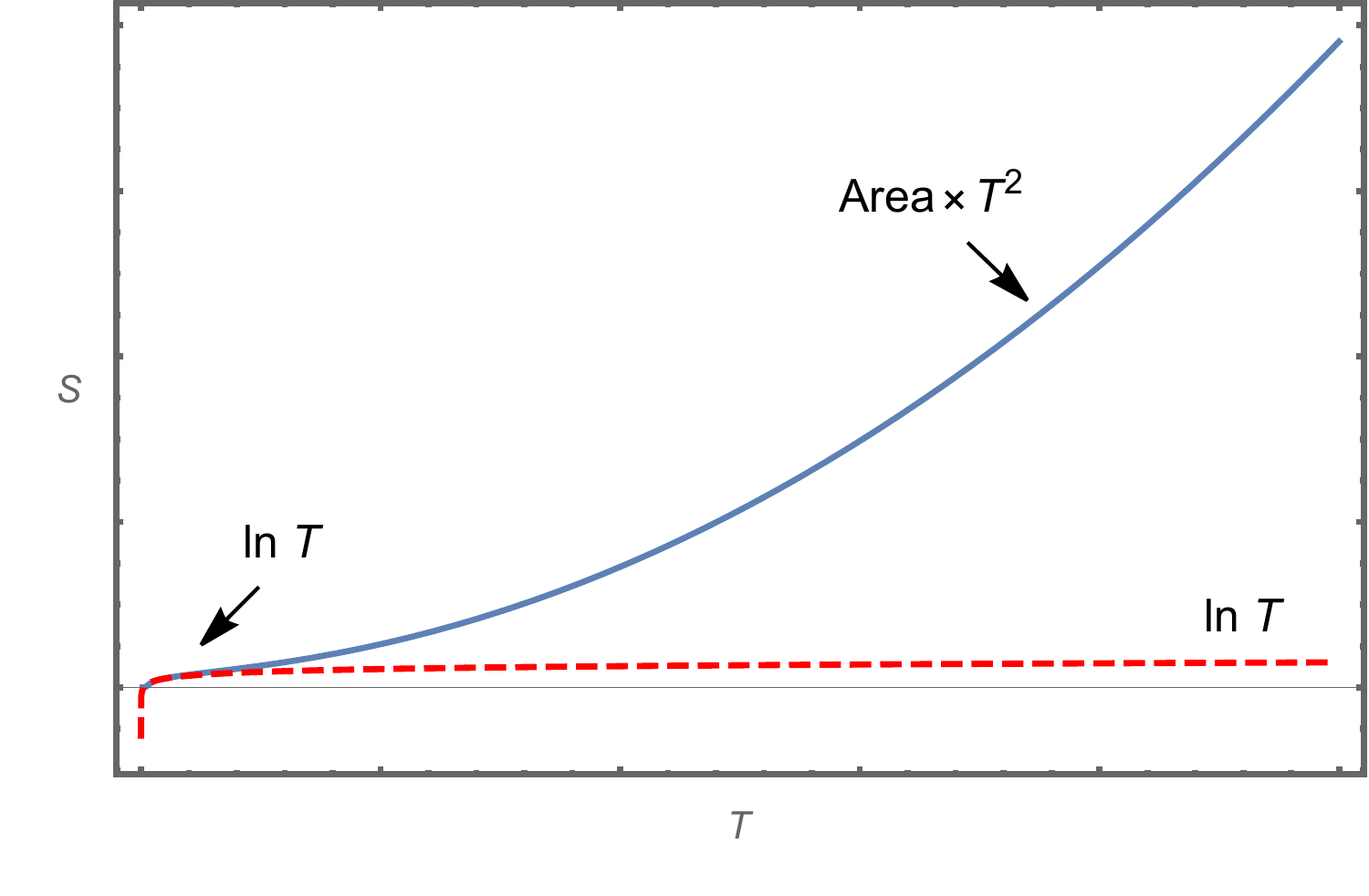}\\
  \includegraphics[width=0.45\textwidth]{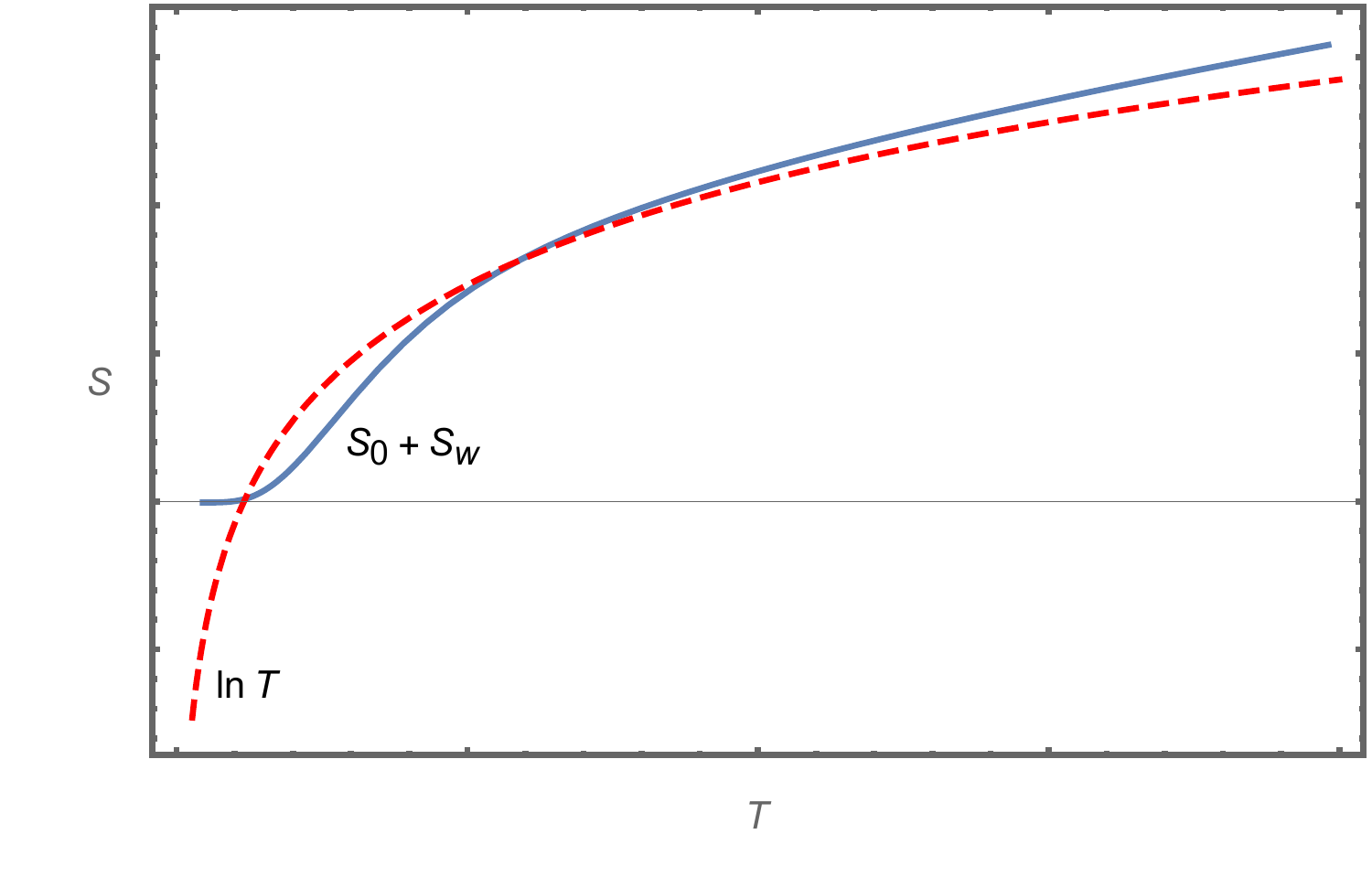}\\
  \caption{A sketch of the entropy of fields $\phi$ and $W$ with varying temperature. At high temperatures, the entropy scales as $ \A\times T^2$. For lower temperatures, the entropy scales as the logarithm of temperature and coupling constant. The second picture is an enlarged version of the low-temperature region. The red dashed line is an auxiliary line showing $\ln T$. As can be seen from the figure, the entropy goes to zero in the super-low temperature limit because the contributions coming from zero modes and winding modes cancel each other. }\label{ST}
\end{figure}

\begin{figure}
  \centering
  \includegraphics[width=0.45\textwidth]{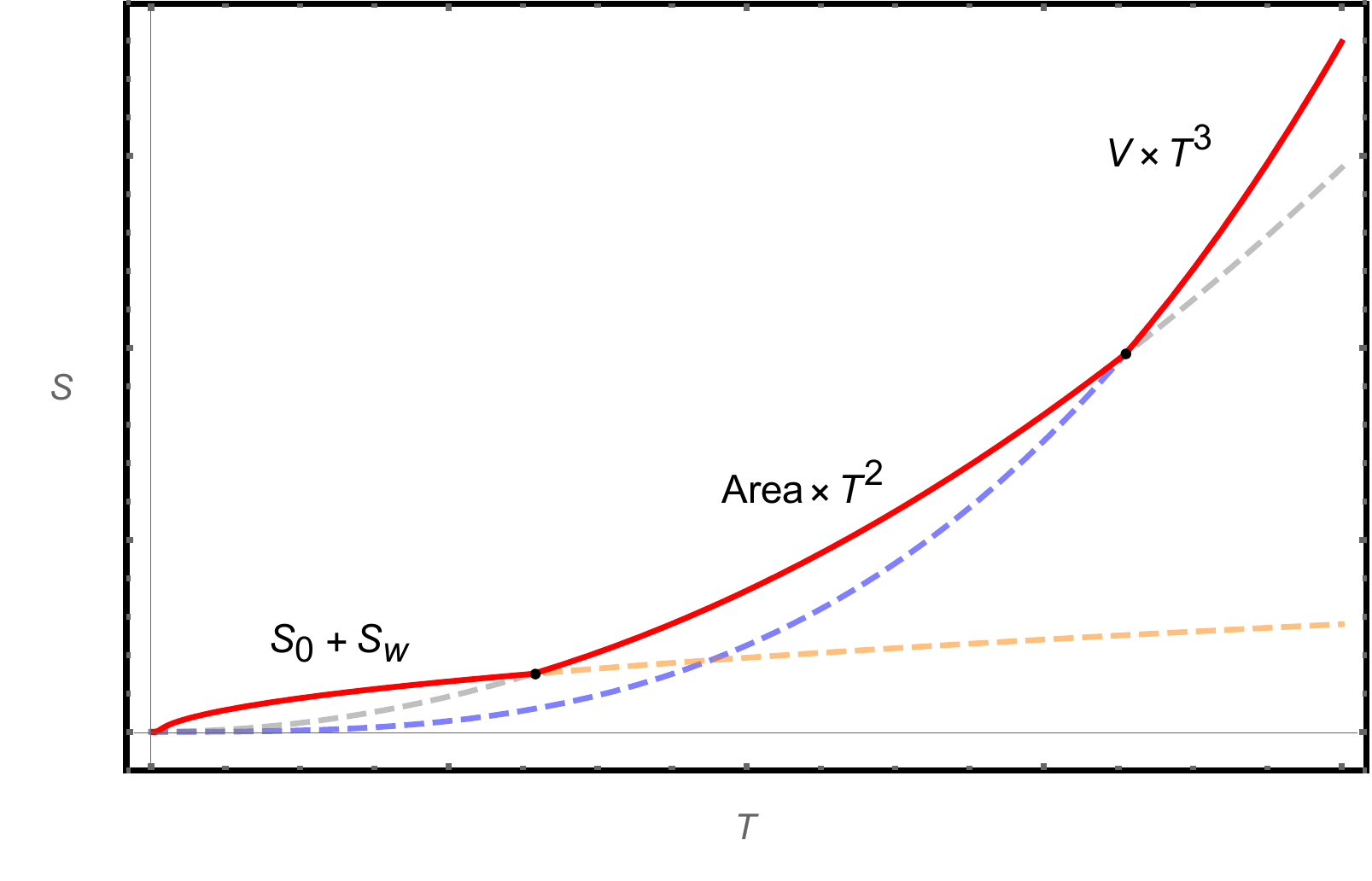}\\
  \caption{A sketch of the entropy of the whole system in different temperature limits. The actual entropy is the sum of different contributions, and the red line demonstrates the dominant contribution. There are two transitions of the dominants shown in the figure. The bulk fluctuation modes always dominate in the high-temperature limit, and the entropy scales as the volume multiplied by temperature cubed. For lower temperature becomes, the area contribution starts to dominate. At super-low temperatures, the fluctuation contribution is not important anymore, and the only contribution is from the constant modes and winding modes of field $W$. A more clear curve of the entropy near the origin is shown in the second panel of figure \ref{ST}. }\label{boxphase}
\end{figure}

~\\
\noindent\textbf{Case I: High temperature limit}~\\~\\
In the high-temperature limit, nothing is special, and we expect to see the usual result of black body radiation in a box because the bulk fluctuation modes, whose entropy is proportional to $T^3$, are the most important contribution.
The partition function $Z_{\hat A}$ and entropy $\mS_{\hat A}$ of bulk modes $\hat A_\mu$ are shown below
\bea
\ln Z_{\hat A} &=& -\frac{1}{8\pi^2} {\beta V}\times{\Lambda^{4}}+ \frac{\pi^2}{45} \frac{V}{\beta^3}\,,\\
\mS_{\hat A} &=& (1-\beta\pd_\beta)\ln Z_{\hat A}= \frac{4\pi^2}{45} {V}T^3\,,\label{ShA}
\eea
which is exactly the blackbody radiation result in a flat box.
In order to avoid confusion with action $S$, we use $\mS_{\hat{A}}$ to denote the corresponding entropy.
When the temperature is high, the contributions from $\phi$ and $W$ proportional to the area of the boundary are small compared with the bulk radiation.
So in the high-temperature limit, the dominant contribution always comes from the bulk fluctuation modes $\hat A_\mu$, which scales as the volume between the two boundaries multiplied by the temperature cubed.

~\\
\noindent\textbf{Case II: Low temperature limit}~\\

For lower temperatures, when we have $L\ll\beta\ll\sqrt{\A}$, the situation starts to change.
In this temperature range, finite $\beta$ means that $1/L$ is very big, and the energy needed to excite high-frequency modes along the $r$ is super high.
Thus, the modes along the $r$ direction are gapped, and we are only left with zero modes along this direction.
Even the zero modes of $\hat A_a$ are killed by the boundary conditions $\hat A_a\big{|}_{\pd \mM}=0$, so we get no contribution from $\hat A_a$ components.
Fortunately, the zero modes of $A_r$, i.e. $\phi$, is a survivor. Moreover, $W$ plays a similar role as $\phi$. The partition function and entropy for $\phi$ and $W$ can be obtained as
\bea
\ln Z &=& \frac{1}{2}\ln\frac{2\pi^2\text{Area}}{e^2\beta L}-\frac{1}{6\pi}{\beta \text{Area}}\cdot{\Lambda^3}+\frac{\zeta(3)}{\pi}\frac{\text{Area}}{\beta^2}+\ln Z_w\,,\nn\\
\mS &=& (1-\beta\pd_\beta)\ln Z\label{Sphi}\\
&=&\frac{3\zeta(3)}{\pi}{\text{Area}}\times T^2+\text{logarithm corrections}\,.\nn
\eea
The entropy of the thermal fluctuation modes along the boundary direction is proportional to the area of the plates times temperature squared. There are extra contributions from constant modes and winding modes if $e^2$ is not very large, which is proportional to the logarithm of temperature and the coupling constant. The logarithm contribution is mainly controlled by the coupling constant $e^2$, which can surpass the area contribution for suitable $e^2$. We will discuss the logarithm contribution later because those terms will be the most important contribution as the temperature is even lower. So in this temperature limit, the entropy of the system mainly comes from the fluctuation modes of $\phi$ and $W$, which is more or less proportional to the area of the plates times the temperature squared.

It is worth noticing that the situation here is similar to the Kuluza-Klein (K-K) reduction along the radius direction. The energy of the K-K tower is proportional to $1/R$, where $R$ is the length scales of extra dimensions. When $R$ is very small, we can only see the zero modes of the K-K tower, and the effective theory is lower-dimensional.

~\\
\noindent\textbf{Case III: Super-low temperature limit}~\\

As the temperature becomes even lower, all the thermal fluctuation contributions proportional to the temperature will not survive.
In the so-called super-low temperature limit, we have $\frac{L\cdot \A}{\beta^3}\gg 1$. The bulk fluctuation modes are already frozen to death in the previous stage, and now it is the turn for the ones of $\phi$ and $W$. The logarithm contributions from the constant modes and winding modes can be written as
\bea
\ln Z = \frac{1}{2}\ln\frac{2\pi^2\text{Area}}{e^2\beta L}+\ln Z_w\,.
\eea
with
\bea\label{lnZw}
\ln Z_w = \left\{
\begin{matrix}
 0~ \, ; &&~~~~ \frac{\A}{2e^2 L \beta}\gg 1 \\
 -\frac{1}{2}\ln[\frac{\A}{\beta L}]+\ln e~. && ~~~~\frac{\A}{2e^2 L \beta}\ll 1
\end{matrix}\right.
\eea
As can be seen from \eqref{lnZw}, the coupling constant $1/e^2$ is a controller of constant modes. For weak coupling, where we have $1/e^2\ll 1$ such that
\be
\frac{1}{e^2}\frac{\text{Area}}{\beta L} \ll 1\,,
\ee
the constant modes contribution of $\phi$ is canceled by the winding modes.
However, in the strong coupling limit
\be
\frac{1}{e^2}\frac{\text{Area}}{\beta L} \gg 1\,,
\ee
we can always see the contribution of the constant modes, which always scales as the logarithm of the coupling constant and temperature.
The corresponding behavior of the entropy of constant modes and winding modes is shown in the second picture of figure \ref{ST}.

As a summary of this section, let us qualitatively illustrate the basic behavior of the entropy corresponding to (\ref{lnZ_phiN}). The bulk fluctuation modes are the most important modes at super-high temperatures, whose entropy should be the blackbody radiation, i.e. $T^3 \times V$. The new ingredient of our story is the modes due to the boundary condition.
At high temperatures, the entropy of the fluctuation modes of $\phi$ and $W$ is proportional to $T^2 \times \A$.
While as temperature goes lower and lower, the fluctuation modes play a less and less important role, and the zero modes and winding modes that contribute as the logarithm of the temperature and $e^2$ start to dominate. However, as the temperature becomes super low, the contribution from $\ln Z_w$ cancels the one from zero modes, and the overall entropy goes to a constant. 
The entropy of the gauge theory with the given boundary condition is illustrated in figure \ref{boxphase}.

\section{Conclusion}
\label{con}

We analyze the partition function of the U(1) gauge field living between two parallel boundaries with boundary condition \eqref{second} in this paper. 
The canonical analysis helps us understand what are the dynamical variables in the phase space (or Hilbert space). We also get the measure of different fields by working out the symplectic form of the theory with the given boundary conditions. 
As shown in figure \ref{cauchy}, the radius coordinate is labeled by $r$ and the transverse coordinates are $x^a$.
Besides the edge modes due to the boundary condition, there are also non-local modes due to the physics interplay between the two boundaries. Those modes are non-local effects, like the Wilson lines stretched between the boundaries, but behave like co-dimension-one fields which are pretty similar to the boundary edge modes. 
So, the physical modes of the theory at hand contain four different parts: bulk fluctuation modes $\hat {A}_{\mu}$, zero longitudinal momentum modes of $A_r$ which is $\phi(x^a)$, boundary stretched Wilson lines $W(x^a)$, constant modes and winding modes.

Putting all of the above modes into the Euclidean path integral, we can work out the partition function of the theory that contains contributions from the four parts.
The bulk fluctuation modes always play the dominant role at very high temperatures, whose entropy scales as the volume of the bulk multiplied by temperature curbed
\be
\mS_{\hat A}\propto \text{Volume}\times T^3\,.
\ee
The modes arising because of the boundary conditions become more and more important as the temperature becomes lower. For lower temperatures, the ratio between the distance of the two boundaries $L$ and inverse temperature $\beta$ becomes small, and the zero modes $\phi$ and the Wilson lines $W$ that behave like boundary scalar fields give out the dominant contributions. The entropy of those modes scales as
\be\label{phiNNN}
\mS_{\phi,W}\propto \A \times T^2\,.
\ee
As the temperature becomes super-low, no fluctuation mode can be seen, and we are left with some constant modes and topological modes contributions. The entropy of those modes is approximately the logarithm of the coupling constant and the temperature. The qualitative behavior of the entropy of different modes is shown in figures \ref{ST} and \ref{boxphase}.

The flat parallel plates case is supposed to serve as a good toy model for the more general situation in curved spacetime.
We would like to see if a similar phenomenon also shows up in the black hole background (or even for the wormhole background). 
The boundary modes and non-local modes should have their counterparts in a black hole system with similar boundary conditions, and the non-local modes might be understood as soft hair of the system.
The phase transitions suggest that there might be similar phenomena for black holes. If not, the special properties of the gravitational system are then encoded in the difference.
We leave the related issues for future research. 

At the end of the paper, let us briefly comment on the first set of boundary conditions, shown in \eqref{Neumann}. It's not ridiculous to add all configurations that respect the boundary condition, including the boundary gauge modes. 
For the U(1) gauge theory, there are bulk on-shell configurations that have one-to-one correspondences with those boundary configurations, and the entropy of those modes can be counted. 
Moreover, the non-gauge invariance of the boundary condition we are mainly interested in this paper naturally captures that kind of physics.
Those boundary pure gauge modes are soft modes because of the vanishing Hamiltonian for the would-be gauge modes. 
Moreover, it was suggested there are possible connections between those modes and soft hair degrees of freedom \cite{Hawking:2016sgy, He:2014laa, Haco:2018ske, Hawking:2016msc, Strominger2018, Cheng:2020vzw, Cheng:2022xyr, Cheng:2022xgm}, Barnich's non-proper degrees of freedom \cite{Barnich:2018zdg, Barnich:2019qex, Alessio:2020lpk, Aggarwal:2022rrp}, and edge modes \cite{Donnelly:2015hxa,Donnelly:2016auv, Blommaert:2018oue, Blommaert:2018rsf, Balachandran:1994vi, Kabat:1994vj, Kabat:2010nm, Seraj:2016jxi, Geiller:2017xad, Seraj:2017rzw, Henneaux:2018gfi}.
More concrete connections between those boundary effects due to boundary conditions are worth further understanding. 

\paragraph{Acknowledgements}
We would like to thank Ankit Aggarwal, Jan de Boer, Diego Hofman, and Pujian Mao for their useful discussions. 
This work is supported by the National Natural Science Foundation of China (NSFC) under Grant No. 11905156 and No. 11935009. 

\appendix

\section{Different temperature limits}
\label{boxtem}

In this appendix, we will evaluate and compare the partition functions of fields $\hat A_\mu$, $\phi$, $W$, and other modes in different temperature limits. The conclusions are summarized in the main context of section \ref{temperature}. Here we would like to provide more details about the calculation. The partition function we intend to evaluate is 
\bea
\ln Z &=& -\sum_{\omega}\sum_{p_r,p_2,p_3} \ln  \left[\beta^2(\omega^2+p_r^2+p_2^2+p_3^2)\right]\nn\\
&~& -\sum_{\omega,p_2,p_3} \ln[\beta^2(\omega^2+p_2^2+p_3^2)]+\frac{1}{2}\ln[\frac{\A}{\beta L}]\nn\\
&~& -\ln e+\ln Z_w\,,\label{B2}
\eea
with
\bea
\ln Z_w= \left\{ \begin{matrix}
 0~\,; &~~~~& \frac{\A}{2e^2 L \beta}\gg 1\\
 -\frac{1}{2}\ln[\frac{\A}{\beta L}]+\ln e~\,. &~~~~& \frac{\A}{2e^2 L \beta}\ll 1
\end{matrix}
\right.
\eea

\subsection{High temperature limit}

First of all, let us take the high-temperature limit $\beta\ll L \ll\sqrt{\A}$. The first task is to evaluate the partition function  $Z_{\hat A}$ for the bulk fluctuation modes. In this temperature limit, we have
 \be
 \frac{V}{\beta^3}=\frac{L\cdot \A}{\beta^3}\gg 1\,,
 \ee
which means that we can write
\bea
\omega &=& \omega_m=\frac{2\pi m}{\beta}\,\\
\sum_{p_r} \sum_{p_2} \sum_{p_3} &=& \frac{V}{(2\pi)^3}\int dp_r dp_2 dp_3\,.
\eea
One can further write the first part in (\ref{B2}) as
\be
\ln Z_{\hat A}=-2V \int \frac{d^3p}{(2\pi)^3}\left[ \frac{1}{2}\beta \omega+\ln (1-e^{-\beta \omega}) \right]\,,
\ee
where we have $\omega=\sqrt{|p_r^2+p_2^2+p_3^2|}$. This is the result for two copies of bosonic fields.
The first part in $\ln Z_{\hat A}$ is ultraviolet (UV) divergent and can be evaluated in the presence of a regulator $\Lambda$. The integrand of the second part is exponentially small as $p$ goes up; thus, the integral is convergent. After introducing UV cutoff $\Lambda$, we have
\be
\ln Z_{\hat A} = -\frac{1}{8\pi^2} {\beta V}\times{\Lambda^{4}}+ \frac{\pi^2}{45} \frac{V}{\beta^3}\,.
\ee
Note that the first part involving UV cutoff $\Lambda$ is a constant in the free energy because the logarithm of the partition function is linear in $\beta$. Therefore, the entropy of those modes can be written as
\be
\mS_{\hat A}=(1-\beta\pd_\beta)\ln Z_{\hat A}= \frac{4\pi^2}{45} \frac{V}{\beta^3}\,.
\ee
For a similar reason, one can show that the partition function for $\phi(x^a)$ and $W$ can be written as
\bea
\ln Z &=& -\frac{1}{6\pi} {\beta \A}\times {\Lambda^3}+\frac{1}{2}\ln\frac{2\pi^2\A}{e^2\beta L}\nn\\
&~& +\frac{\zeta(3)}{\pi}\frac{\A}{\beta^2}+\ln Z_w\,.
\eea
The winding modes contribution $Z_w$ shown above depends on the value of coupling constant $e^2$. For the case
\be
\frac{1}{e^2}\frac{\A}{\beta L}\gg 1\,,
\ee
we have
\be
Z_w=\sum_n e^{-\frac{\text{Area}}{2e^2 \beta L}(2\pi n)^2}\approx e^{-\frac{\text{Area}}{2e^2 \beta L}(2\pi n)^2}\big{|}_{n=0}=1\,,
\ee
thus $\ln Z_w=0$.
However, when $e^2$ is big enough such that
\be
\frac{1}{e^2}\frac{\A}{\beta L}\ll 1\,,
\ee
the coefficient inside of the exponential function is very small, and we can change the sum into an integral. Thus we have
\be
Z_w\approx\int dn~ e^{-\frac{\A}{2e^2 \beta L}(2\pi n)^2}=\left(\frac{2\pi \A}{e^2\beta L}\right)^{-1/2}\,,
\ee
the logarithm of which can be written as
\be
\ln Z_w=-\frac{1}{2}\ln \frac{2\pi\text{Area}}{e^2\beta L}\,.
\ee
So when $e^2$ is large enough, the contribution of the constant modes $\ln Z_0$ can be canceled. Nevertheless, this does not matter because the constant modes contribution of $\phi$ is always much smaller than the fluctuation modes contributions in this temperature limit. The statistical entropy of the fluctuation modes of $\phi$ and $W$ can be computed as
\bea
\mS &=& (1-\beta\pd_\beta)\ln Z= \frac{3\zeta(3)}{\pi}\frac{\text{Area}}{\beta^2}\label{boxSF}\,.
\eea
Compared with the volume contribution, all the area and logarithm contributions are not going to be important.
As shown in equation (\ref{ShA}), the most important contribution always comes from the bulk fluctuation modes $\hat A_\mu$, which scales as the volume times the temperature cubed.

\subsection{Low temperature limit}

High temperature is boring because we can only see the bulk fluctuation modes. As the temperature goes lower, when we have $L\ll \beta\ll\sqrt{\text{Area}}$, interesting phenomena due to the boundary condition start to show up.

First, let us look at the bulk fluctuation modes $\ln Z_{\hat A}$.
In this limit, the distance between the two plates is very small compared to the inverse temperature $\beta$. Assuming finite temperature, we have $\omega_m=2\pi m/\beta$.
Small $L$ implies the high-frequency modes along the $r$ direction are gapped, and we would only see zero modes along the $r$ direction. $\hat A_\mu$ vanish on the boundary, so the zero modes of  $\hat A_\mu$ along the $r$ direction is killed by the boundary conditions. The only surviving zero modes are the zero modes of $A_r$ namely $\phi$, which will be discussed separately. In the low-temperature limit and also in the super-low temperature limit, we will never see any contribution from bulk fluctuation modes anymore. So we can conclude that the entropy from the bulk fluctuation modes is
\be
\mS_{\hat A}=0\,.
\ee



However, the zero modes $\phi$ survived and the partition functions for $\phi$ and $W$ are not changed. We still have
\bea
&~&\ln Z \\ &=&
\frac{1}{2}\ln\frac{2\pi^2\text{Area}}{e^2\beta L}-\frac{1}{6\pi}{\beta \text{Area}}\cdot{\Lambda^3}+\frac{\zeta(3)}{\pi}\frac{\text{Area}}{\beta^2}+\ln Z_w\,.\nn
\eea
For the case $\frac{1}{e^2}\frac{\text{Area}}{\beta L}\gg 1\,$, we have $\ln Z_w=0$, and the overall partition function can be written as
\bea
\ln Z &=& \frac{1}{2}\ln\frac{2\pi^2\text{Area}}{e^2\beta L}-\frac{1}{6\pi}{\beta \text{Area}}\cdot{\Lambda^3}+\frac{\zeta(3)}{\pi}\frac{\text{Area}}{\beta^2}\,.
\eea
The corresponding entropy can be calculated as
\bea
\mS 
\approx \frac{3\zeta(3)}{\pi}\frac{\text{Area}}{\beta^2}
+\frac{1}{2}\ln\frac{2\pi^2\text{Area}}{e^2\beta L}+\frac{1}{2}\,.
\eea
For the case where $e^2$ is very large $\frac{1}{e^2}\frac{\text{Area}}{\beta L}\ll 1$,
the contribution of the constant modes $\ln Z_0$ is canceled by the winding modes contribution, we only left with fluctuation modes contribution. The overall entropy is
\be
\mS=(1-\beta\pd_\beta)\ln Z=\frac{3\zeta(3)}{\pi}\frac{\text{Area}}{\beta^2}\,.
\ee
The entropy of the system now scales the area times the temperature squared.

\subsection{Super-low temperature limit}

As the temperature becomes even lower, we have $L\ll\sqrt{\text{Area}}\ll\beta$, which is the low-temperature limit. In this temperature limit, not only the contribution from $\hat A_\mu$ can be ignored, the fluctuation modes of $\phi$ and $W$ are not important at all.
As for the constant modes and winding modes, we have
\bea
\ln Z = \frac{1}{2}\ln\frac{2\pi^2\text{Area}}{e^2\beta L}+\ln Z_w\,.
\eea
And in the limit $\frac{1}{e^2}\frac{\text{Area}}{\beta L}\gg 1$, we have $\ln Z_w=0$. The overall entropy can be written as
\be
\mS=\frac{1}{2}\ln\frac{2\pi^2\text{Area}}{e^2\beta L}+\frac{1}{2}\,.
\ee
Whereas in the limit $\frac{1}{e^2}\frac{\text{Area}}{\beta L}\ll 1$, we have
\be
\ln Z_w=-\frac{1}{2}\ln \frac{2\pi\text{Area}}{e^2\beta L}
\ee
which cancels the constant modes and the overall entropy tends to be a small constant.

\providecommand{\href}[2]{#2}\begingroup\raggedright\endgroup


\end{document}